\documentclass[twocolumn,showpacs,preprintnumbers,amsmath,amssymb]{revtex4}

\usepackage{graphicx}
\usepackage[colorlinks,dvipdfm]{hyperref}

\usepackage{subfigure}

\usepackage{dcolumn}
\usepackage{bm}
\usepackage{amsthm}
\usepackage{amsmath}
\usepackage{extarrows}
\usepackage{multirow}
\usepackage{longtable}
\usepackage{rotating}

\begin{document}

\preprint{APS/123-QED}

\title{Grover search with smaller oracles}

\author{Dan Li$^{1}$, }
\affiliation{ $^{1}$ College of Computer Science and Technology, Nanjing University of Aeronautics and Astronautics, Nanjing, China\\
        }

\date{\today}

\begin{abstract}

Grover search is one of the most important quantum algorithms. In this paper, we consider a kind of search that the conditions of satisfaction $T$ can be rewritten as $T=T_1\bigcap T_2$. Then we present a new Grover search with smaller oracles. The time complexity of this algorithm $O(\frac{\pi}{4}\sqrt{\frac{N}{b\lambda}}+\frac{\pi}{4}\sqrt{\frac{b}{\tau}})$, which is smaller than the  time complexity of original Grover search, i.e. $O(\frac{\pi}{4}\sqrt{\frac{N}{M}})$.

\end{abstract}

%
\maketitle

\section{Introduction}
\label{sec:level1}
Search on a unordered database is one of the NP-hard problems. The classical way  to execute the exhaustive search is by querying each item in the database of $N$ items by a oracle to identify the solution. In the worst case, the total number of queries to the oracle is $N-1$.

Grover search is one of the most important quantum algorithm, which is presented by Grover in Ref.\cite{G001,G002}. Grover's algorithm can find one target item with oracle complexity $O(\sqrt{N})$, which quadratically outperforms the classical algorithm.

Grover partial search is presented from the view that only some part of bits of the database are interested \cite{PG001,PG002}. The authors use a local Grover operator to make the partial search  easier. Then Choi, Zhang and Korepin consider quantum partial search of a database with several target items \cite{PGeven,PGNeven}. Then Zhang and Korepin discuss how to optimise the Grover's algorithm from the view of depth \cite{PGwithOpt}.

By borrowing the idea of local Grover operator, we present the Grover search algorithm with smaller oracles in this paper. By consider the  conditions of satisfaction $T$ as $T=T_1\bigcap T_2$, the time complexity of the new Grover search is smaller than the original Grover's search.

The paper is structured as follows. In Sect.\ref{sec:level2}, we review the Grover search algorithm. In Sect.\ref{sec:level3},  the Grover search algorithm with smaller oracles and its quantum circuit are presented. And the time complexity of this algorithm is discussed.  Finally, a short conclusion is given in Sect.\ref{sec:level5}.

\section{Grover search algorithm}
\label{sec:level2}

The quantum search algorithm consists of repeated application of a quantum subroutine, know as the Grover iteration, which we denote $G$.  The Grover iteration, whose quantum circuit is illustrated in Fig1XXXXX, may be broken up into two steps:

\begin{enumerate}
  \item Apply the oracle $O_T$;
  \item Apply the Grover operator $D_1$.
\end{enumerate}

$O_T$ is a quantum oracle with the ability to recognize solutions to the search problem. The action of the oracle may be written as:

\begin{equation}\label{E201}
  O_T=I-2\sum_{x} |x\rangle\langle x|
\end{equation} which in fact has the effect $|x\rangle\stackrel{O_T}{\longrightarrow}-|x\rangle$ for all target items.

The Grover operator $D_1$ is
\begin{equation}\label{E202}
  D_1=2|\psi_1\rangle\langle \psi_1|-I,
\end{equation} which is the inversion about mean operation. $|\psi_1\rangle$ is the equal superposition of all items in the database.

Suppose $N$ is the size of database, $M$ is the number of targets. The initial state is $|\psi_1\rangle$. Let $CI(x)$ denote the integer closest to the real number $x$. Then the number of Grover iteration is
\begin{equation}\label{E203}
  R=CI(\frac{arccos\sqrt{M/N}}{arccos(1-2\frac{M}{N})}),
\end{equation} which is $O(\sqrt{N/M})$.

\section{The algorithm of Grover search with smaller oracles}
\label{sec:level3}

By borrowing the idea of  local Grover operator, we present the algorithm of Grover search with smaller oracles.

Suppose the conditions of satisfaction $T$ can be rewritten as $T=T_1\bigcap T_2$. $T_1$ is the condition of satisfaction of the first $log(k)$ qubits, while $T_2$ is the condition of satisfaction of all qubits, i.e. $T$ or part of them. Based on the above limitations, $T\subset T_1$ and $T \subseteq T_2$ .

A database of $N$ items is divided into $k$ blocks  with $N=bk$. Here $b$ is the number of items in each block.

The idea of this algorithm is shown in Fig.2XXX.

Firstly, consider all items  that satisfy $T_1$ as target items,  after the global Grover iterations, amplitudes of the target blocks which satisfy $T_1$ get higher while amplitudes of non-target blocks are close to 0.

Secondly, consider items  that satisfy $T_2$ as target items, then after the local Grover iterations, amplitudes of  the items in the target blocks get higher more. Because the total of amplitudes of a non-target block is close to 0,  after the local Grover iterations, amplitudes of target items in non-target blocks are still close to 0.


Blocks that satisfy the condition $T_1$ are denoted by $YY_i$, whose number is $\lambda_Y$. And the set of target items, i.e. satisfy $T_2$, in these blocks are denoted by $AY_i$, whose size is $\tau_i$, while the set of non-target items in these blocks are denoted by $XY_i$, whose size is $b-\tau_i$.

Blocks which include items that satisfy the condition $T_2$, but do not satisfy the condition $T_1$, are denoted by $NY_i$, whose number is $\lambda_N$. And the set of items which satisfy $T_2$ in these blocks are denoted by $AN_i$, whose size is $\omega_i$, while the complementary set of them in each block is denoted by $XN_i$, whose size is $b-\omega_i$.

Blocks that do not satisfy the condition $T_1$  or $T_2$ are denoted by $B$. The number of these blocks is $k-\lambda_Y-\lambda_N$

Here we define some states.

\begin{eqnarray}\label{E}
  |AY_i\rangle &=&\frac{1}{\sqrt{\tau_i}}\sum_{x\in AY_i} |x\rangle \\
  |XY_i\rangle &=&\frac{1}{\sqrt{b-\tau_i}}\sum_{x\in XY_i}  |x\rangle  \\
  |YY\rangle &=&\frac{1}{\sqrt{b\lambda_Y}}\sum_{x\in \bigcup YY_i}  |x\rangle  \\
           &=& \sum_i \sqrt{\frac{\tau_i}{b\lambda_Y}} |AY_i\rangle+\sum_i \sqrt{\frac{b-\tau_i}{b\lambda_Y}} |XY_i\rangle  \\
   |AN_i\rangle &=&\frac{1}{\sqrt{\omega_i}}\sum_{x\in AN_i} |x\rangle \\
  |XN_i\rangle &=&\frac{1}{\sqrt{b-\omega_i}}\sum_{x\in XN_i}  |x\rangle  \\
  |NY\rangle &=&\frac{1}{\sqrt{b\lambda_N}}\sum_{x\in \bigcup NY_i}  |x\rangle  \\
           &=& \sum_i \sqrt{\frac{\omega_i}{b\lambda_N}} |AN_i\rangle+\sum_i \sqrt{\frac{b-\omega_i}{b\lambda_N}} |XN_i\rangle  \\
    |NN\rangle &=&\frac{1}{\sqrt{N-b(\lambda_Y+\lambda_N)}}\sum_{x\in B}  |x\rangle \\
   |B\rangle &=&\frac{1}{\sqrt{N-b\lambda_Y}}\sum_{x\in NY\bigcup NN}  |x\rangle \\
\end{eqnarray}

\begin{figure}\label{Fig2}
 \begin{center}
  \subfigure[]{
\label{Fa}
\includegraphics[width=8cm]{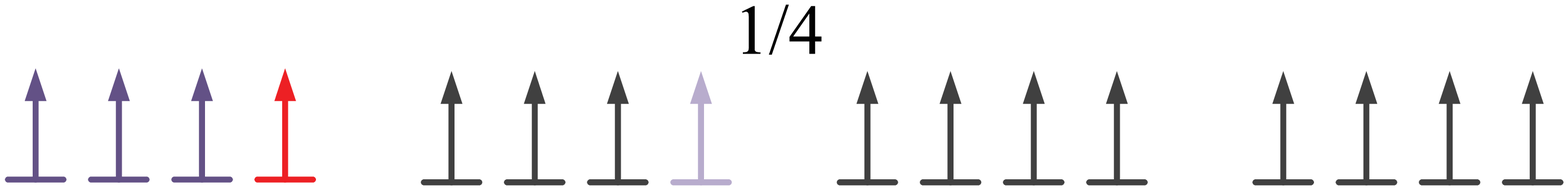}} \\
  \subfigure[]{
  \label{Fb}
 \includegraphics[width=8cm]{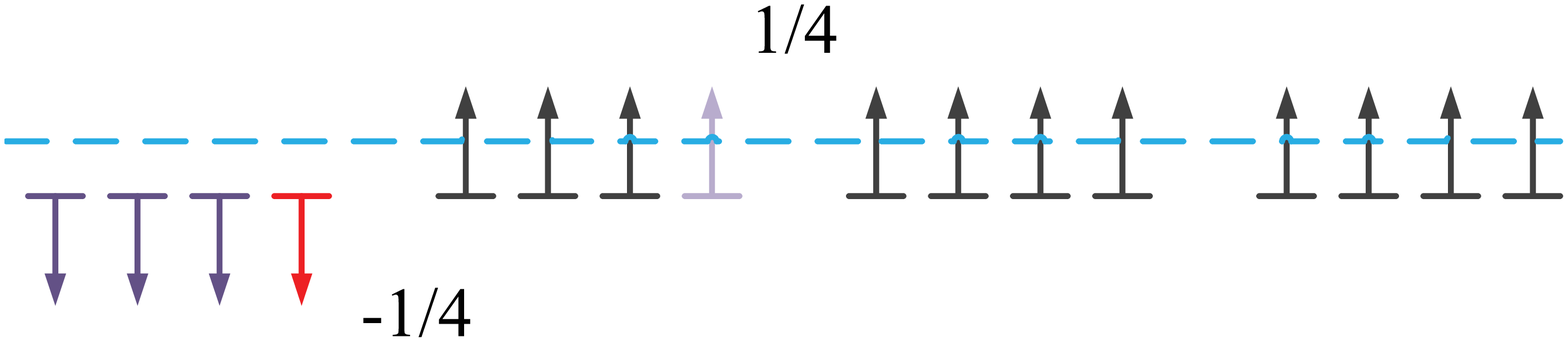}} \\
  \subfigure[]{
  \label{Fc}
\includegraphics[width=8cm]{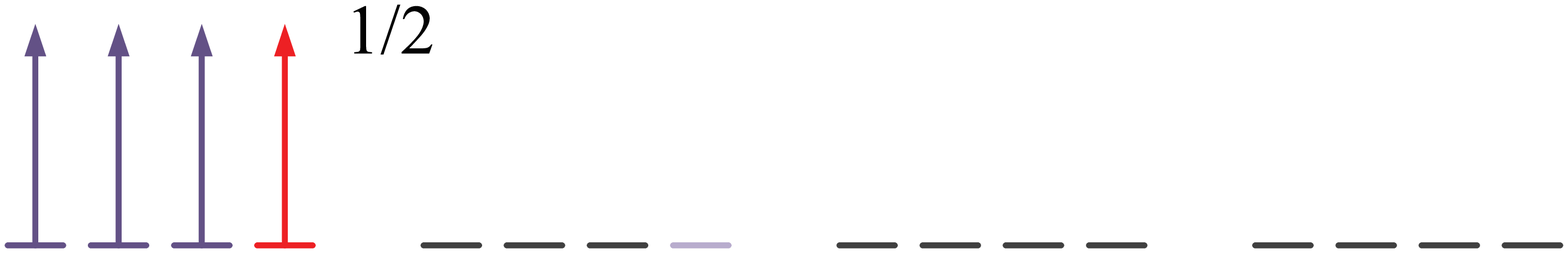}}\\
\subfigure[]{
\label{Fd}
\includegraphics[width=8cm]{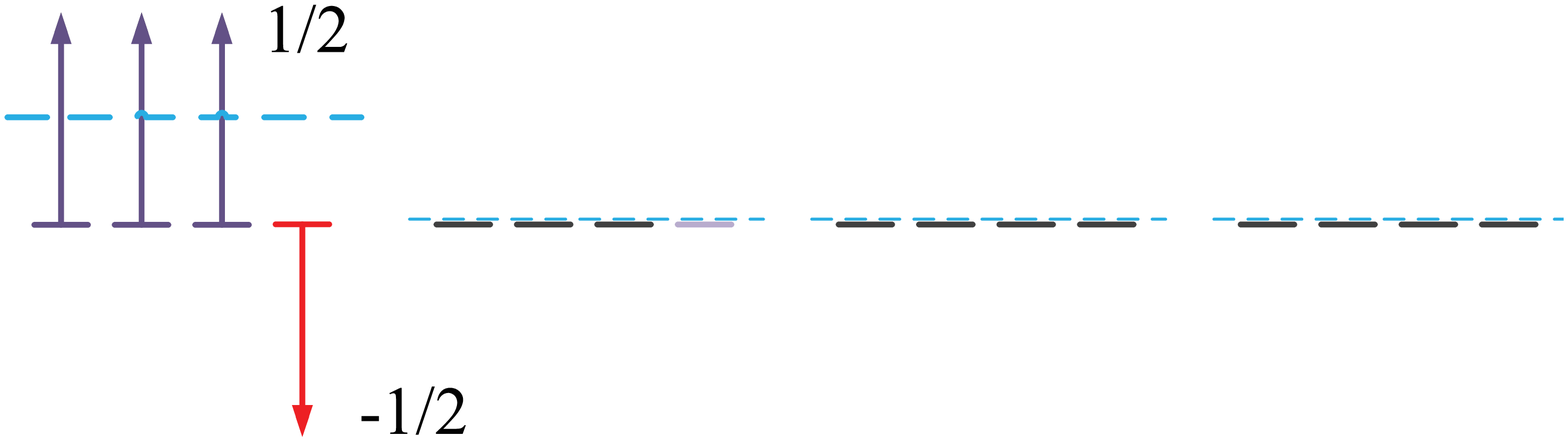}} \\
\subfigure[]{
\label{Fe}
\includegraphics[width=8cm]{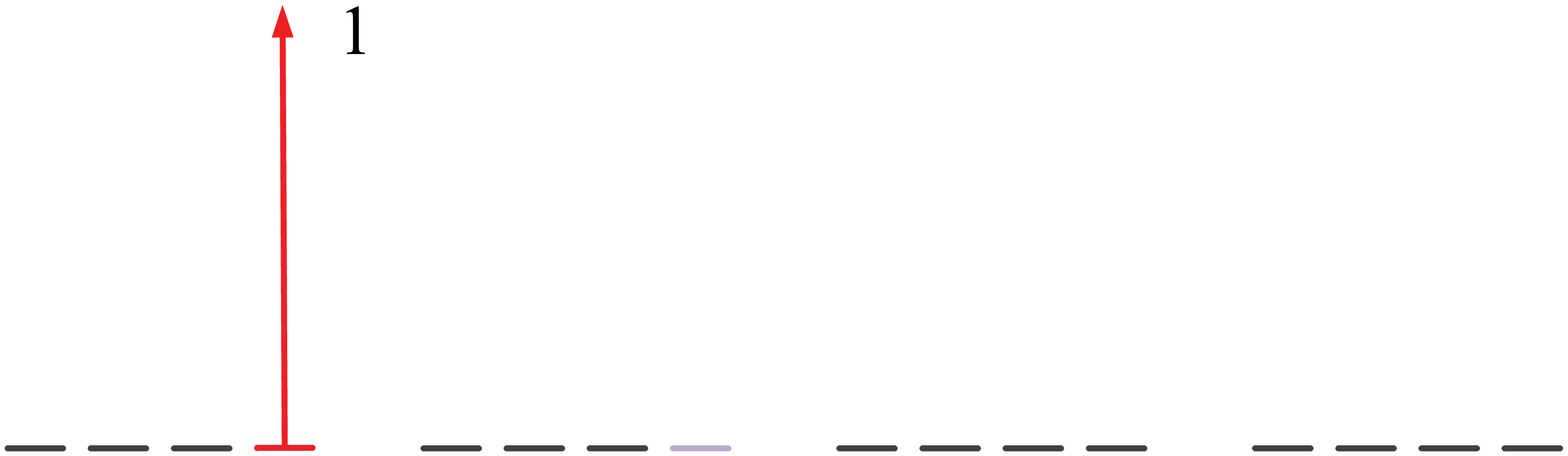}} \\
  \end{center}
\renewcommand{\figurename}{Fig.}
\caption{The two Grover iterations $G_1$ and $G_2$. Items, No.1-4, satisfy $T_1$, Items, No.4,8, satisfy $T_2$.}
\end{figure}

\subsection*{Step1: $j_1$ the Grover iteration $G_1$;}

The Grover iteration $G_1$ is defined as
\begin{eqnarray}\label{E}
  G_1=D_1\ast O_{T1},
\end{eqnarray}where

\begin{eqnarray}\label{E}
  O_{T1}=I-2\sum_{x\in \bigcup YY_i} |x\rangle\langle x|\\
  D_1=2|\psi_1\rangle\langle \psi_1|-I
\end{eqnarray}

The effect of $G_1$ on $|YY\rangle$ and $|B\rangle$ is

\begin{eqnarray}\label{E}
  \widehat{G}_1=\left(
                  \begin{array}{cc}
                    cos(2\theta) & sin(2\theta) \\
                    -sin(2\theta) & cos(2\theta) \\
                  \end{array}
                \right)
\end{eqnarray}where $sin^2[\theta]=\frac{b\lambda_Y}{N}$

Therefore, after step 1, the state of the system is

\begin{eqnarray}\label{E}
 && \widehat{G}_1^{j_1} |\psi_1\rangle=a_1|YY\rangle+ a_2 |B\rangle   \nonumber\\
 && =\sum_i \sqrt{\frac{\tau_i}{b\lambda_Y}} a_1 |AY_i\rangle+
 \sum_i \sqrt{\frac{b-\tau_i}{b\lambda_Y}} a_1 |XY_i\rangle  \nonumber \\
 && +\sum_i \sqrt{\frac{\omega_i}{N-b\lambda_Y}} a_2 |AN_i\rangle   \nonumber\\
 && +\sum_i \sqrt{\frac{b-\omega_i}{N-b\lambda_Y}} a_2 |XN_i\rangle  \nonumber \\
  && +\sqrt{\frac{N-b\lambda_Y-b\lambda_N}{N-b\lambda_Y}}a_2|NN\rangle
\end{eqnarray} where $a_1=sin[(2j_1+1)\theta]$, $a_2=cos[(2j_1+1)\theta]$.

\subsection*{Step2: $j_2$ the Grover iteration $G_2$;}

The Grover iteration $G_2$ is defined as
\begin{eqnarray}\label{E}
  G_2=D_2\ast O_{T2},
\end{eqnarray}where

\begin{eqnarray}\label{E}
  O_{T2}=I-2\sum_{x\in \bigcup AY_i\bigcup AN_i} |x\rangle\langle x|\\
  D_2=I_k\bigotimes (2|\psi_2\rangle\langle \psi_2|-I)   \\
  |\psi_2\rangle=\frac{1}{\sqrt{b}} \sum_{x_2\in\{1,\cdots b\}} |x_2\rangle
\end{eqnarray}

The effect of $G_2$ on $|AY_i\rangle$ $|XY_i\rangle$ $|AN_i\rangle$ $|XN_i\rangle$ and $|B\rangle$ is

\begin{eqnarray}\label{E}
  \widehat{G}_2=\left(
                  \begin{array}{ccccc}
                    cos(2\theta_{Yi}) & sin(2\theta_{Yi}) & \ & \ & \ \\
                   -sin(2\theta_{Yi}) & cos(2\theta_{Yi}) & \ & \ & \  \\
                   \ & \ & cos(2\theta_{Ni}) & sin(2\theta_{Ni}) & \  \\
                    \ & \ &  -sin(2\theta_{Ni}) & cos(2\theta_{Ni})  & \  \\
                     \  & \  &  \  & \   & 1 \\
                  \end{array}
                \right)
\end{eqnarray}where $sin^2[\theta_{Yi}]=\frac{\tau_i}{b}$,  $sin^2[\theta_{Ni}]=\frac{\omega_i}{b}$.

Therefore, after step 2, the state of the system is

\begin{eqnarray}\label{E}
  &&\widehat{G}_2^{j_2}\widehat{G}_1^{j_1} |\psi_1\rangle=
    \sum_i (\sqrt{\frac{\tau_i}{b\lambda_Y}} a_1bi_2+ \sqrt{\frac{b-\tau_i}{b\lambda_Y}} a_1bi_1 |AY_i\rangle  \nonumber \\
  &&+\sum_i (-\sqrt{\frac{\tau_i}{b\lambda_Y}} a_1bi_1+ \sqrt{\frac{b-\tau_i}{b\lambda_Y}} a_1bi_2 |XY_i\rangle    \nonumber\\
  &&   \sum_i (\sqrt{\frac{\omega_i}{N-b\lambda_Y}} a_2ci_2+ \sqrt{\frac{b-\omega_i}{N-b\lambda_Y}} a_2ci_1 |AY_i\rangle  \nonumber \\
  &&+\sum_i (-\sqrt{\frac{\omega_i}{N-b\lambda_Y}} a_2ci_1+ \sqrt{\frac{b-\omega_i}{N-b\lambda_Y}} a_2ci_2 |XY_i\rangle    \nonumber\\
&& +\sqrt{\frac{N-b\lambda_Y-b\lambda_N}{N-b\lambda_Y}}a_2|NN\rangle
\end{eqnarray} where $bi_1=sin[2j_2\theta_Yi]$, $bi_2=cos[2j_2\theta_Yi]$, $ci_1=sin[2j_2\theta_Ni]$, $ci_2=cos[2j_2\theta_Ni]$.

\subsection*{Estimating the number of Grover iterations}

The first step of this algorithm is to magnify the amplitude of blocks $YY_i$, i.e. the amplitude of $|YY\rangle$. Therefore, the optimal number of Grover iteration $G_1$ is $j_1=CI(\frac{arccos\sqrt{\frac{b\lambda_Y}{N}}}{arccos(1-2\frac{b\lambda_Y}{N})})$.

The second step of this algorithm is to magnify the amplitude of  items that satisfy $T_2$ in each block. In non-target blocks $NY_i$, amplitudes of all items are close to 0. So the number of Grover iteration $G_2$ do not affect greatly the amplitudes. In target blocks, items which satisfy $T_2$ are items satisfy $T$. The optimal number of Grover iteration $G_2$ for each target block $YY_i$ is $CI(\frac{arccos\sqrt{\frac{\tau_i}{b}}}{arccos(1-2\frac{\tau_i}{b})})$. Due to the number of target items in each target blocks may not same, in order to minimize the number of Grover iteration $G_2$, the optimal number of Grover iteration $G_2$ is $j_2=CI(\frac{arccos\sqrt{\frac{\tau}{b}}}{arccos(1-2\frac{\tau}{b})})$, where $\tau=max(\tau_i)$.

For Grover search, the number of Grover iteration is  $O(\frac{\pi}{4}\sqrt{\frac{N}{M}})$. For the Grover search with smaller oracles in this paper, the time complexity is  $O(\frac{\pi}{4}\sqrt{\frac{N}{b\lambda}}+\frac{\pi}{4}\sqrt{\frac{b}{\tau}})$, which is smaller than $O(\frac{\pi}{4}\sqrt{\frac{N}{M}})$.

\section{Summary}
\label{sec:level5}

By borrowing the idea of  local Grover operator, we present a new Grover search algorithm with smaller oracles. This algorithm is suitable for search questions whose  conditions of satisfaction $T$ can be rewritten as $T=T_1\bigcap T_2$, where $T_1$ and $T_2$ are conditions of satisfaction for part of qubits.

The algorithm is divided into two parts: global Grover iteration, local Grover iteration. Global Grover iteration magnify the amplitude of items that satisfy $T_1$, while local Grover iteration magnify the amplitude of items that satisfy $T_2$.

On  one hand, the time complexity of this algorithm is  $O(\frac{\pi}{4}\sqrt{\frac{N}{b\lambda}}+\frac{\pi}{4}\sqrt{\frac{b}{\tau}})$, which is smaller than the  time complexity of original Grover search, i.e. $O(\frac{\pi}{4}\sqrt{\frac{N}{M}})$. On the other hand, this algorithm needs more to evaluate the number of target items that satisfy $T_1$ and $T_2$.

In conclusion, this algorithm is not as general as the original Grover search, but in specific situation, it is more fast, and only need smaller oracles.

\begin{acknowledgments}
This work is supported by NSFC (Grant Nos. 61701229, 61702367，61901218), Natural Science Foundation of Jiangsu Province, China (Grant Nos. BK20170802，BK20190407), China Postdoctoral Science Foundation funded Project (Grant Nos. 2018M630557, 2018T110499), Jiangsu Planned Projects for Postdoctoral Research Funds (Grant No. 1701139B), the Open Fund of the State Key Laboratory of Cryptology (Grant  No. MMKFKT201914).

\end{acknowledgments}

\bibliography{basename of .bib file}
\bibliography{apssamp}

\end{document}